\RequirePackage{fix-cm}
\documentclass{svjour3}                     
\smartqed  
\usepackage{graphicx}
\usepackage{natbib}
\usepackage{mathptmx}  
\begin{document}

\title{Effect of migration in a diffusion model for template coexistence in protocells
}


\author{Jos\'e F. Fontanari         \and
  Maurizio Serva      
}

\authorrunning{Fontanari and Serva} 

\institute{ J. F. Fontanari  \at
               Instituto de F\'{\i}sica de S\~ao Carlos,
  Universidade de S\~ao Paulo,
  Caixa Postal 369, 13560-970 S\~ao Carlos SP, Brazil\\
              \email{fontanari@ifsc.usp.br}
           \and
         M. Serva \at
Departamento de Biof\'{\i}sica e Farmacologia,
Universidade Federal do Rio Grande do Norte,
59072-970 Natal, RN, Brazil   \\
On leave of absence from
Dipartimento di Ingegneria e Scienze 
dell'Informazione e Matematica, Universit\`a 
dell'Aquila, I-67010 Coppito, L'Aquila, Italy.      
}

\date{Received: date / Accepted: date}

\maketitle

\begin{abstract}
The compartmentalization of  distinct templates in protocells  and the exchange  of templates between them (migration)
are key elements of a modern scenario for prebiotic evolution. Here we use  the diffusion approximation of population genetics to 
study analytically the steady-state properties of such prebiotic  scenario.
The coexistence of distinct template types inside a protocell is
achieved by  a selective pressure  at the protocell level (group selection) favoring protocells with a mixed 
template composition.   In the degenerate case,
where the templates have the same replication rate, we find that  a vanishingly small migration rate suffices to
eliminate the segregation effect of random drift and so to promote coexistence. In the non-degenerate case, a small migration rate
greatly boosts coexistence as compared with the situation where there is no migration. However, increase of the migration rate
beyond a critical value
leads to the complete dominance of the more efficient template type (homogeneous regime). In this case,
we find  a continuous phase transition  separating the homogeneous and the coexistence  regimes, 
with the order parameter vanishing linearly with the distance to the transition point.
\end{abstract}

\section{Introduction}\label{sec:intro}

The coexistence of competing  selfish individuals is an ubiquitous  issue in the study of systems  described by 
the modern Darwinian paradigm, known as the ``Evolutionary Synthesis''  \citep{Maynard_95,Mayr_01,Mayr_02,Mayr_04,Nowak_04}. 
In the context of prebiotic or chemical evolution, this matter surfaced
with the observation by \citet{Eigen_71} that, due to the  finite fidelity of replication, 
the information content of a single self-replicating macromolecule (a template for short)
is far too few to permit the coding  of macromolecules with any functional complexity.
A way out of this difficulty, so-called the information crisis of prebiotic evolution,
 is to assume the information is distributed among a number of distinct template types and
enforce cyclic cooperative interactions among them -- the hypercycle --  to guarantee coexistence \citep{Eigen_78,Eigen_80}.

Alternatively, coexistence between distinct template types can be achieved by confining the templates in  packages
or protocells and  requiring that the survival or the reproduction chances of a
protocell be dependent on its template composition \citep{Bresch_80,Niesert_81,Szat_87}.  The study of
this two-level selection problem can be carried out by introducing  minor changes on the
mathematical models developed to address the efficiency of group selection to maintain an altruistic trait
\citep{Eshel_72,Aoki_82,Donato_97}.
In particular,  in a recent paper we have used a diffusion model of group selection  \citep{Kimura_83}
to study analytically the conditions for the coexistence of two template types which differ on their replication rates
\citep{Fontanari_13}. However,  that  study   left out a crucial characteristic of the primitive protocell populations, namely,
the elevated exchange flux of templates among protocells, known as lateral or horizontal gene transfer. In fact,
the acceptance of  the operation of this process in the early history of microbial life has wiped out completely the
familiar Darwinian notion of a universal  ancestor \citep{Woese_98,Doolittle_00}.

Here we model the process  of template swapping among protocells by the classic migration process of 
Wright's island model \citep{Wright_51}. We find that introduction of migration renders the evolutionary
process ergodic 
in the sense that the steady state does not depend  on the initial set-up of the population. In addition, migration 
allows  a steady-state solution corresponding to protocells carrying  both template types (coexistence regime)
or a solution where  the more efficient  template type is fixed in all protocells 
(homogeneous regime). There is a smooth transition between these two regimes   provided that the two template types  
exhibit distinct replication rates. In the degenerate case, where the template types  have identical replication efficiencies, only the  coexistence
regime is stable. This contrasts with the results obtained  in the absence of migration, for which there is a non-ergodic segregation regime  characterized
by a mixture of two types of protocells, each type  carrying solely one of the template types \citep{Fontanari_13}.

The remainder of the paper is organized as follows. In Sec.\ \ref{sec:model} we describe the  three
evolutionary processes -- template competition, migration and intercell competition --  that comprise the
dynamics of our two-level selection model, and derive the partial differential equation that governs the time evolution of
the fraction of protocells carrying a given template composition. 
Sec.\  \ref{sec:Steady} is devoted to the numerical and analytical study of the steady-state 
solutions of that equation. In particular, our numerical approach relies on the
interpretation of the steady-state ordinary differential equation  as an eigenvalue  problem 
whose eigenvalue corresponds to the  mean group selection pressure.
 Our concluding remarks are presented in Sec.\ \ref{sec:Conc}. In Appendix A
we present the analytical calculation of the probability that a template type fixates
in a given protocell in the non-ergodic segregation regime for the case migration is
not allowed. This calculation generalizes that presented in \citet{Fontanari_13} by
taking into account the different replication efficiencies of  the template types.

\section{The model}\label{sec:model}

Following \cite{Kimura_83}, we consider a hypothetical population divided into an infinite number of
competing protocells, each of which containing exactly $N$ templates. There are two types of templates
which differ only by their replication efficiency: type 1 templates have a selective disadvantage $s$ relative to type 2
templates, where $s \geq 0$ is a parameter on the order of $1/N$. More pointedly, type 1 templates are assigned fitness $1-s$  and
type 2 templates 
fitness $1$.  In addition, we assume that 
$N$ is large enough so that the frequency of  type 1 templates  within a protocell, denoted by  $x$,
can be viewed as a continuous variable in the interval $ \left [ 0,1 \right ]$. Of course, the frequency of
type 2 templates within the same protocell is $1-x$.
The population   is
described by  the fraction of protocells $\phi \left ( x, t \right ) \Delta x$ whose frequency of type 1 templates lies in the range $\left ( x , x + \Delta x \right )$  at time $t$. Our goal is to determine  how the probability density $\phi \left ( x, t \right )$ is affected
by the three  evolutionary processes: individual template competition within a protocell, migration of templates between protocells
and competition between protocells.  

The template competition process  within each  protocell takes place according to the rules of the standard 
Wright-Fisher model of population genetics \citep{Crow_70}.  In particular, assuming that a protocell contains 
$j$ type 1 templates and  $N-j$ type 2 templates, the probability that  there will be exactly $i$ type 1 templates after
template competition  is given by the  Wright-Fisher process
\begin{equation}\label{rij}
r_{ij} = \left( {\begin{array}{*{20}c} N \\ i \\ \end{array}} \right) w_j^i \left ( 1 - w_j \right )^{N-i},
\end{equation}
where $w_j = j \left ( 1 - s \right )/\left ( N - j s  \right )$ is the relative fitness of the subpopulation of
type 1 templates in the protocell. 
To determine how this process affects the probability density $\phi \left ( x, t \right )$ we resort to the   diffusion 
approximation of population genetics \citep{Crow_70}, which consists essentially on the calculation of
the jump moments $\left \langle \left ( x' - x \right ) \right \rangle_r$ and $\left \langle \left ( x' - x \right )^2 \right \rangle_r$
where $x = j/N$ and $x'= i/N$  are the frequencies of type 1 templates before and after  template competition, respectively. Here
$\langle \ldots \rangle_r$ stands for an average using the transition probability $r_{ij}$. These moments
contribute to the  drift  and the diffusion  terms of a forward Kolmogorov-like equation for $\phi \left ( x, t \right )$. More pointedly,
direct evaluation of the jump moments to first order in  $1/N$ using the transition probability (\ref{rij}) yields 
\begin{equation}\label{d1r}
\left \langle \left ( x' - x \right ) \right \rangle_r = w_j - x \approx - s x \left ( 1 - x \right )  
\end{equation}
and 
\begin{equation}\label{d2r}
\left \langle \left ( x' - x \right )^2 \right \rangle_r =  \frac{1}{N} w_j \left ( 1 - w_j \right ) + \left ( w_j - x \right )^2  \approx \frac{1}{N}  x \left ( 1 - x \right ),  
\end{equation}
where  we have used that the fitness disadvantage $s$ of the type 1 templates  is on the order of $1/N$.

Migration follows Wright's island model \citep{Wright_51} that posits  that  $J$  templates of each protocell are replaced by migrants
in the time interval $\Delta t$  and that the frequency of type 1 templates among the migrants is  equal to the average frequency
of type 1 templates in the entire protocell population, i.e., $\bar{x} = \int_0^1 x \phi \left ( x, t \right ) dx$.  The probability
that a  protocell with $j$ type 1 templates ($x = j/N$) becomes a protocell with $i$ type 1 templates ($x' = i/N$) due to the migration process is then
\citep{Aoki_82}
\begin{equation}
m_{ij} = \sum_{k=k_l}^{k_u} \frac{ \left( {\begin{array}{c} j \\ k \\ \end{array}} \right) \left( {\begin{array}{c} N -j\\ J-k \\ \end{array}} \right)}
{\left( {\begin{array}{c} N \\ J \\ \end{array}} \right)}  \left( {\begin{array}{c} J \\ i -j + k\\ \end{array}} \right) 
\bar{x}^{i-j+k} \left ( 1 - \bar{x} \right )^{J-i+j-k},
\end{equation}
where $k_l = \max \left (j-i,0,J-N+j\right)$ and $k_u = \min \left ( j,J-i+j,J \right )$. 
Here the hyper-geometric component  yields the probability that exactly 
$k$ type 1 templates and $J - k$ type 2 templates are eliminated from the protocell to make room for the $J$ migrants, whereas the binomial part yields the probability that
there are exactly $i-j+k$ type 1 templates among the $J$ migrants. After migration the number of type 1 templates in the protocell is given by the sum of 
the type 1 templates originally in the protocell  $\left ( j-k \right )$  and the number of type 1 templates among the migrants $\left ( i-j+k \right )$.
The first two jump moments are given by
\begin{equation}\label{d1m}
\left \langle \left ( x' - x \right ) \right \rangle_m = m \left ( \bar{x} - x \right ) 
\end{equation}
and 
\begin{equation}\label{d2m}
\left \langle \left ( x' - x \right )^2 \right \rangle_m=  \frac{m}{N} \bar{x} \left ( 1 - \bar{x} \right ) + m^2 \left ( \bar{x}- x \right )^2  
+ \frac{m \left ( 1 - m \right )}{N-1}  x \left (1 - x \right ),
\end{equation}
where $\langle \ldots \rangle_m$ stands for an average using the transition probability $m_{ij}$ and 
 $m = J/N$ is the fraction of the protocell population that is replaced by migrants. Assuming that $m$ is on the  order of
$1/N$, i.e., that the number of migrants $J$  remains finite and limited when $N$ grows large,  we can
neglect the second jump moment which is $O \left ( 1/N^2 \right )$. 

Finally, the competition between protocells is taken into account as follows. Denoting  by $c \left ( x \right )$ the
selection coefficient of a protocell with a fraction $x$ of type 1 templates  we have 
\begin{equation}
\phi \left ( x , t + \Delta t \right ) = \left [ \phi \left ( x , t \right ) + c \left ( x \right )  \phi \left ( x , t \right )  \Delta t \right ] \zeta,
\end{equation}
where $\zeta$ is such that $\int_0^1  \phi \left ( x , t + \Delta t \right ) dx  =1$, i.e, 
$\zeta = 1/\left [ 1 +   \bar{c}\left( t \right )  \Delta t \right ]$ with
\begin{equation}\label{cbar}
\bar{c} \left ( t \right ) = \int_0^1 c \left ( x \right ) \phi \left ( x , t  \right )  dx .
\end{equation}
Taking the limit $\Delta t \to 0$ we obtain the change in the fraction of protocells due to intercell
selection, $\Delta \phi = \left [ c \left ( x \right ) - \bar{c} \left ( t \right ) \right ]  \phi \left ( x , t  \right ) \Delta t$.

Combining the changes in $\phi$  due to the three processes described above and introducing the rescaled variables $\tau = t/2N$, $S= 2Ns \geq 0$, $M=2Nm \geq 0$ and $C \left ( x \right ) =2N c \left ( x \right ) \geq 0$ we obtain \citep{Kimura_83}
\begin{equation}\label{eqphi}
\frac{\partial}{\partial \tau} \phi \left (x,\tau \right )  = \frac{\partial^2}{\partial x^2} \left[ x \left (1-x \right )\phi \left (x, \tau \right ) \right]
-\frac{\partial}{\partial x} \left[ b \left (x, \tau \right ) \, \phi \left (x,\tau \right ) \right]
+ \left  [ C \left ( x \right )-\bar{C} \left (\tau \right ) \right ] \phi  \left (x, \tau \right ),
\end{equation}
where  
\begin{equation}\label{drift}
b \left (x,\tau \right ) = -S x \left (1-x \right ) - M \left [ x- \bar{x} \left (\tau \right ) \right ]
\end{equation}
is the drift term,
\begin{equation}
\bar{x} \left (\tau \right )=\int_0^1 x \, \phi \left (x,\tau \right ) \, dx 
\end{equation}
is the mean number of type 1 templates in the 
protocell population,  
and 
\begin{equation}
\bar{C} \left (\tau \right ) =  \int_0^1 C \left (x \right ) \, \phi \left (x,\tau \right )\, dx 
\end{equation}
is the mean group selection pressure.  The constraint $\int_0^1 \phi \left (x,\tau \right ) \, dx = 1$ 
holds for all times $\tau$.

We note that whereas the linear forward Kolmogorov equation is the standard output in the case  of random drift and  individual selection
\citep{Crow_70}, eq.\ (\ref{eqphi}) is nonlinear because of the presence of $\bar{x}\left (\tau \right )$ and $\bar{C} \left (\tau \right )$, which are
associated to migration and group selection. In addition, the  singularities (if any) of the solution of eq.\ (\ref{eqphi})  must be integrable
so as to guarantee that it is normalizable for all times.

Kimura's choice for the intercell selection coefficient, $C(x) \propto x$, aimed at  exploring 
the efficiency of group selection to maintain  an altruistic character -- the type 1 template  in that case --  
which has a selective disadvantage $s$ relative  to  its competitor  but 
whose presence would boost  the protocell reproduction rate, which increases linearly with the frequency of altruists
inside it.
We refer the reader to  \cite{Ogura_87}  for a rigorous analysis of
the  linear intercell selection model introduced by \cite{Kimura_83} and to \cite{FS_13} for the
analysis of the nonlinear variant of Kimura's model. 
Here we consider the coexistence problem instead, 
which is more burdensome to group selection than the altruistic version, since the fixation of a template
type  through the effect of random drift, regardless of its selective advantage or disadvantage,
acts against coexistence \citep{Fontanari_06}. According to  the so-called metabolic model of template cooperation
\citep{Bresch_80,Niesert_81,Szat_87,Czaran_00,Silvestre_08}, in order to favor coexistence we choose the intercell selection coefficient
\begin{equation}\label{presc}
C \left ( x \right ) = C x \left ( 1 - x \right )
\end{equation}
which is maximum for well-balanced protocells at which $x=1/2$. Here $C$ is a parameter on the order of $1$
that measures the intensity of the group selection pressure towards coexistence. The idea behind 
eq.\  (\ref{presc}) is that
 the two functional template types coded for a small piece of a modular enzyme which then promoted  protocell replication  \citep{Manrubia_07}.
Since the hookup of the
replicase requires products from the two template types, its production rate is proportional to the concentration of the rare  type, hence
the requirement that $c \left ( x \right )$ is maximized by well-balanced protocells. 

The model has three  parameters, namely, $S$ that measures the selective disadvantage of type 1 templates  in
the within cell competition process,
$M$ that measures the strength of migration, and $C$ that measures the strength of the group selection pressure towards
template coexistence. The scale of these parameters is given by the coefficient of the diffusion term which is set to $1$ in 
eq.\  (\ref{eqphi}).

\section{The steady-state solutions}\label{sec:Steady}

The steady-state  protocell probability density  $ \phi = \phi \left ( x \right ) = \lim_{\tau  \to \infty} \phi \left (x,\tau \right )$ satisfies 
\begin{equation}\label{stat1}
 \frac{d^2}{d x^2} \left[ x \left (1-x \right )\phi  \right]
+\frac{d}{d x} \left[ S x \left (1-x \right ) \, \phi  + M \left ( x- \bar{x}  \right ) \, \phi  \right]
+ \left  [C  x \left (1-x \right )-\bar{C}  \right ] \phi  = 0
\end{equation}
with $\bar{x} = \lim_{\tau  \to \infty} \bar{x} \left (\tau \right )$,  $\bar{C} = \lim_{\tau  \to \infty} \bar{C} \left (\tau \right )$, 
and $\int_0^1  \phi  \left ( x \right ) dx = 1$.

For $M > 0$, eq.\ (\ref{stat1})  is satisfied both by $\phi = \delta \left ( x \right )$ and    $\phi = \delta \left ( x-1 \right )$, and it may also  be
satisfied by a regular function $\phi = \phi_r \left ( x \right )$.  By a regular solution of eq.\ (\ref{stat1}) we intend a non-vanishing continuous function
$\phi_r$ in the interval $\left [0,1 \right ]$ which is of class $C^2$ in 
$\left (0,1 \right )$ where it satisfies (\ref{stat1}).
Since $\phi_r$ is a probability density we additionally restrict to
normalizable functions, i.e. functions such that 
$\int_0^1 \phi_r \left (x \right ) dx$ is finite.
We note that in the absence of the coexistence pressure $C=0$, the regular solution is missing 
\citep{Crow_70}. However, one can easily verify  that the migration term prohibits solutions which are combinations of 
the three possibilities (i.e., the  deltas at $x=0$ and $x=1$ and the regular solution), since in that case
eq.\  (\ref{stat1}) would be violated in one of the two extremes, $x=0$ or $x=1$. Clearly, each possibility corresponds to  protocell  populations with 
distinct characteristics. In particular,  $\phi = \delta \left ( x \right )$ describes a population composed of type 2 templates only, 
$\phi = \delta \left ( x-1 \right )$ a population of type 1 templates only, and  $\phi = \phi_r \left ( x \right )$ describes the desired
situation   where the
different templates cohabit  a same protocell.

It is instructive to note that if a regular solution exists,  then integration of eq.\ (\ref{stat1}) over the interval $\left [ 0, 1 \right ]$
yields
\begin{equation}
\left. \frac{d}{dx} \left ( x \phi_r \right ) - M \bar{x} \phi_r \right |_{x=0} = 0 
\end{equation}
and
\begin{equation}
\left. \frac{d}{dx} \left [ \left (1- x \right )\phi_r \right ] - M \left ( 1 - \bar{x} \right ) \phi_r \right |_{x=1} = 0
\end{equation}
which imply that for $x$ close to $0$ one has $\phi_r \sim x^{M \bar{x} -1}$, whereas for $x$ close to $1$ one has
$\phi_r \sim \left ( 1 - x \right )^{M \left ( 1-\bar{x}\right ) -1}$. Hence, in spite of the fact that $\phi_r$ describes a regime of
coexistence, this coexistence can be very unbalanced in the sense that the majority of the protocells may be populated by
essentially a single template type.  This unbalance is typical in the case $M<1$.

%
\subsection{Numerical analysis}
%

The steady-state solutions  of a diffusion model of intergroup selection for the maintenance of an altruistic trait were obtained numerically by
\cite{Kimura_83} in the simple case of a linear group selection pressure $c \left ( x \right ) \propto x$, i.e., $\bar{C} \propto \bar{x}$. In that case 
eq.\ (\ref{stat1}) exhibits  only one non-local term 
   and a straightforward self-consistent iterative approach yields the correct solution.
In our case such a direct approach is doomed to failure, as it will become clear below. 

Following \cite{Kimura_83} we write the regular solution of  eq.\ (\ref{stat1}) in the form 
$ \phi_r \left ( x \right )= \kappa  \phi_0 \left ( x \right )  \psi \left ( x \right ) $ where $\phi_0$ is the solution 
 in  the absence of group selection ($C=0$) and for fixed $\bar{x} \neq 0,1$, namely,
\begin{equation}\label{phi0}
\phi_0 = \exp \left ( -Sx  \right ) x^{ M \bar{x} - 1} \left ( 1 - x \right ) ^{ M(1-\bar{x}) -1} ,
\end{equation}
and $\kappa$ is  the normalization constant. Hence the equation for $\psi$ reads
\begin{equation}\label{psi}
x \left  (1-x \right ) \frac{d^2\psi }{d x^2} -  \left  [ Sx \left(1-x \right ) + M \left(x-\bar{x} \right ) \right ] \frac{d \psi }{d x} + 
C x \left  (1-x \right ) \psi =  \bar{C}  \psi,
\end{equation}
which, as already pointed out, for fixed $\bar{x}$ can be viewed as an eigenvalue problem without boundary conditions that can be
solved by requiring the regularity of $\psi \left ( x \right )$ in $\left [ 0,1 \right ]$  only \citep{Chalub_09}.
In addition, according to the expected behavior of $\phi_r$ in the vicinities of $x=0$ and $x=1$ we can guarantee
that $\psi$ is bounded at these extreme values. Of course, the attempt to solve  eq.\ (\ref{psi}) numerically 
for an arbitrary  value of $\bar{C}$ using, say the Runge-Kutta algorithm,   results in divergences at the extremes, 
which ruins  any self-consistent iterative approach to solve this equation.

Next we define $\psi = \exp(y)$ and get the following nonlinear equation
\begin{equation}\label{eqy}
x \left  (1-x \right ) \left [ y'' + \left ( y' \right )^2 \right ] -  \left  [ Sx \left (1-x \right ) + M \left (x-\bar{x} \right ) \right ] y'
+ C x \left  (1-x \right )  =  \bar{C},   
\end{equation}
where the primes indicate derivatives with respect to $x$.  This is really a first order equation
for $ z \equiv y'$,
\begin{equation}\label{z}
x \left  (1-x \right ) \left [ z' + z^2 \right ] -  \left  [ Sx \left (1-x \right ) + M \left (x-\bar{x} \right ) \right ] z
+  C x \left  (1-x \right )  =  \bar{C},  
\end{equation}
with 
\begin{equation}\label{z0}
z (0) = \frac{\bar{C}}{M \bar{x}}
\end{equation}
and
\begin{equation}\label{z1}
z (1) = - \frac{\bar{C}}{M \left ( 1 - \bar{x} \right )} .
\end{equation} 
At this stage the problem is ready for a numerical approach. For fixed $\bar{x}$ and $\bar{C}$ we solve 
eq.\ (\ref{z}) by propagating the Runge-Kutta algorithm from $x=0$ to $x=1$ using the initial condition 
(\ref{z0}).
%
%
Of course, the choice of an arbitrary value of $\bar{C}$ will not satisfy the boundary condition (\ref{z1}) so we 
adjust $\bar{C}$ in order that condition is satisfied. This is essentially an application of the well-known shooting method to solve boundary values
problems \citep{Press_92}. Once this is achieved, we have solved the problem  for a fixed $\bar{x}$.
Explicitly, $y$ is obtained from
\begin{equation}
y = \int_0^x  z (\xi ) d\xi ,
\end{equation}
where we have defined $y \left ( 0 \right ) = 0$ (hence $\psi \left ( 0 \right ) = 1$). This choice is inconsequential since the physical quantities
are given by ratios of integrals involving $\psi = e^y$. In fact, we can then calculate $\bar{x}$,
\begin{equation}
\bar{x} = \frac{\int_0^1 x \phi_0 (x) e^y dx}{\int_0^1  \phi_0 (x) e^y dx} ,
\end{equation}
return to eq. (\ref{z}) and repeat the process until  we reach the convergence for $\bar{x}$. In particular, we 
assume that convergence occurs whenever the change in $\bar{x}$ is less than $10^{-6}$ in two consecutive iteration steps. This  iterative 
scheme is extremely efficient since it involves the numerical solution of a single first-order ordinary differential
equation  and the iteration over a single quantity only, namely $\bar{x}$.

\begin{figure}
\includegraphics{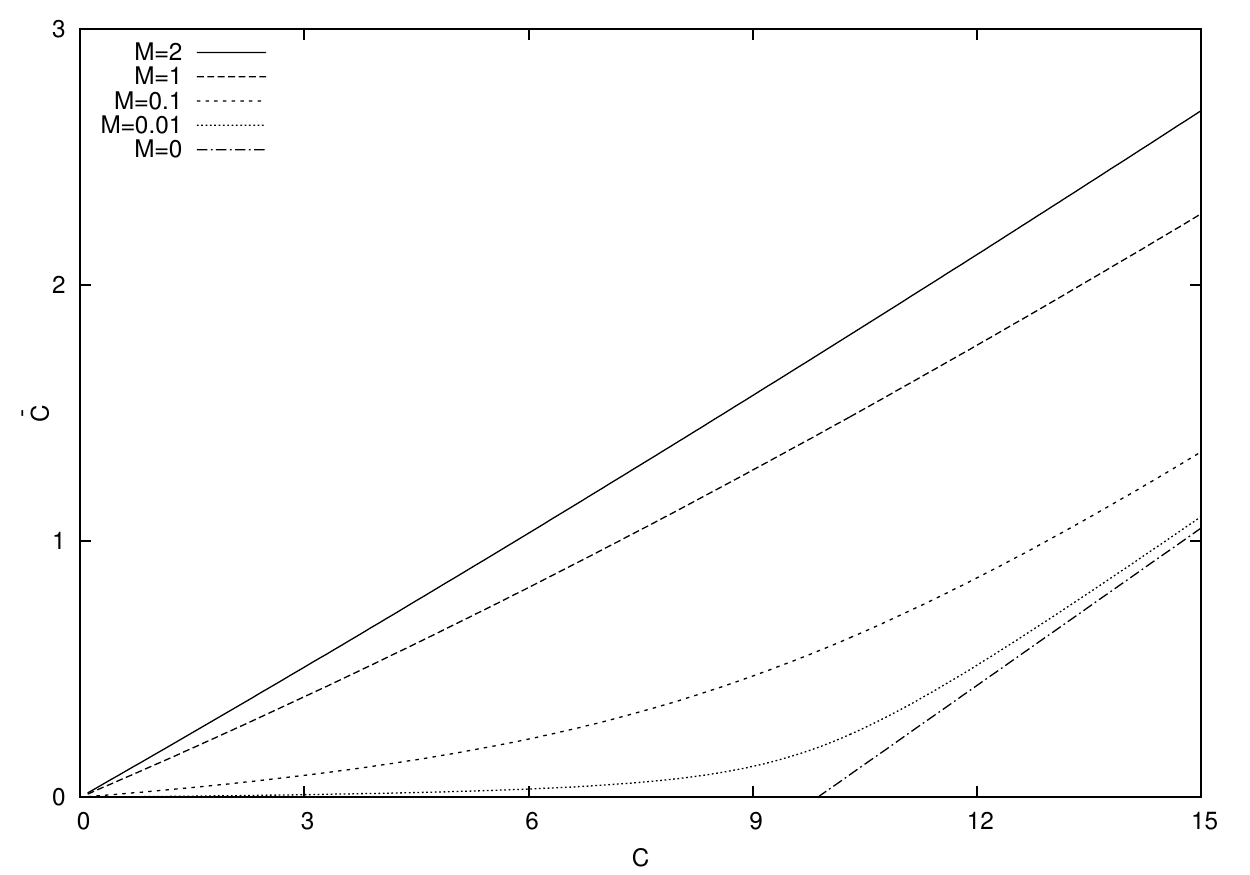}
\caption{Eigenvalue  $\bar{C}$  of the eigenvalue problem (\ref{psi}) as function of  the coexistence pressure  $C$ for 
the degenerate case $S=0$ and migration parameter $M$  as indicated in the figure.  A phase transition takes place at $C_c = \pi^2$ in the case $M=0$. }
\label{fig:1}
\end{figure}
\begin{figure}
\includegraphics{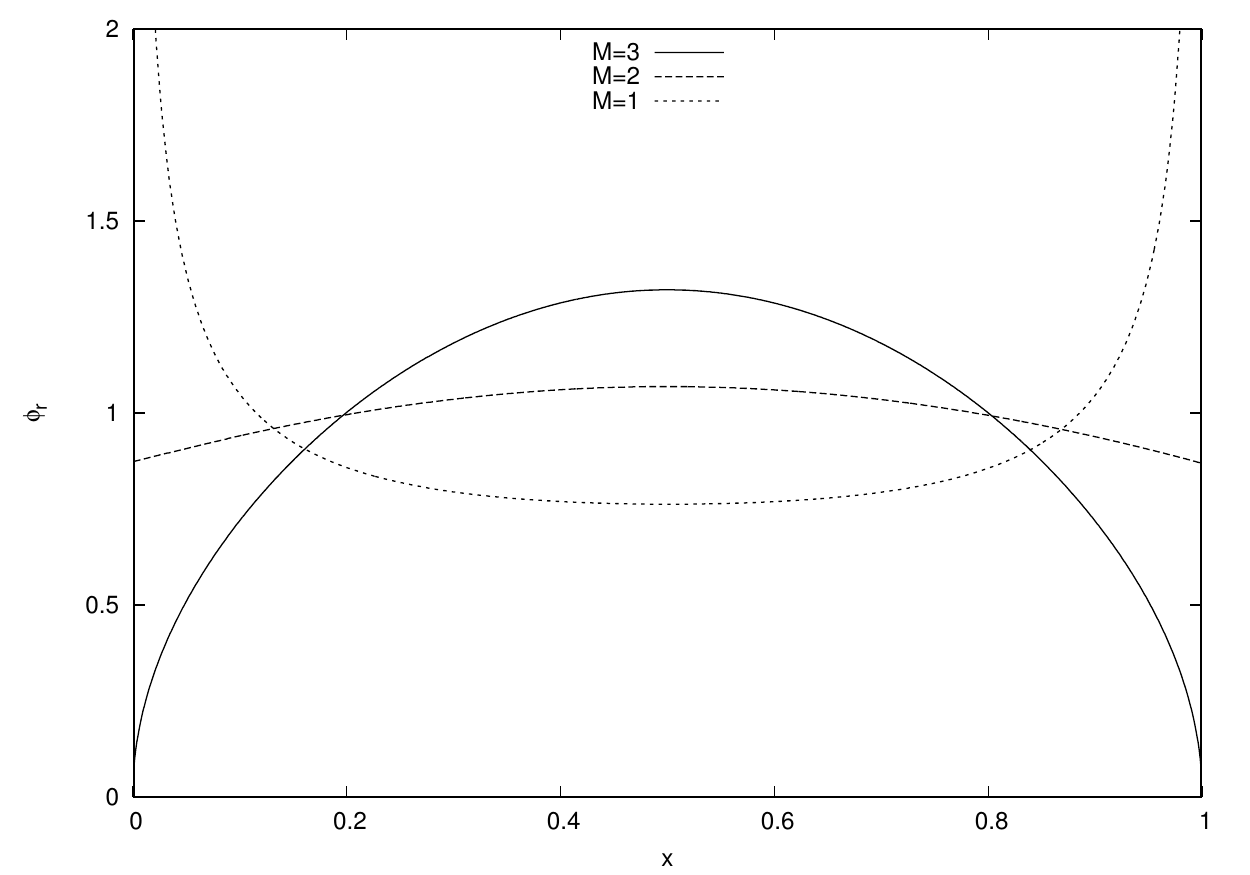}
\caption{Regular normalized steady-state solution  $\phi_r$ giving the  proportion of protocells  that contain a fraction $x$ of type 1 templates
for the degenerate case  $S=0$, coexistence pressure $C=5$ and 
migration rates $M=3, 2, 1$ as indicated in the figure.}
\label{fig:2}
\end{figure}

In Fig.\ \ref{fig:1} we show the dependence of the eigenvalue $\bar{C}$ on the coexistence pressure parameter $C$ in the 
degenerate case   $S=0$ and for a variety 
of values of the migration parameter. In this case, the symmetry of eqs.\  (\ref{stat1}) and (\ref{psi}) with respect to the interchange of $x$ 
and $1-x$  yields $\bar{x} = 1/2$ regardless of the values of $M$ and $C$.  This is illustrated in Fig.\ \ref{fig:2} where the regular solution 
$\phi_r$ is shown for representative values of the migration parameter.
Interestingly, the phase transition between the coexistence 
($\bar{C} > 0$)
and the  segregation ($\bar{C} = 0$) phases that takes place at $C= \pi^2$ for $M=0$  and $S=0$ \citep{Fontanari_13} disappears altogether
when the process of migration is included in  the model. The segregation phase, which is characterized by
a well-balanced mixture of protocells composed of either type 1 or type 2 templates, is eliminated in this case. Hence, in the degenerate case
where there is no  selective advantage at the template level  ($S=0$), migration promotes coexistence (see Fig.\ \ref{fig:2}). 

\begin{figure}
\includegraphics{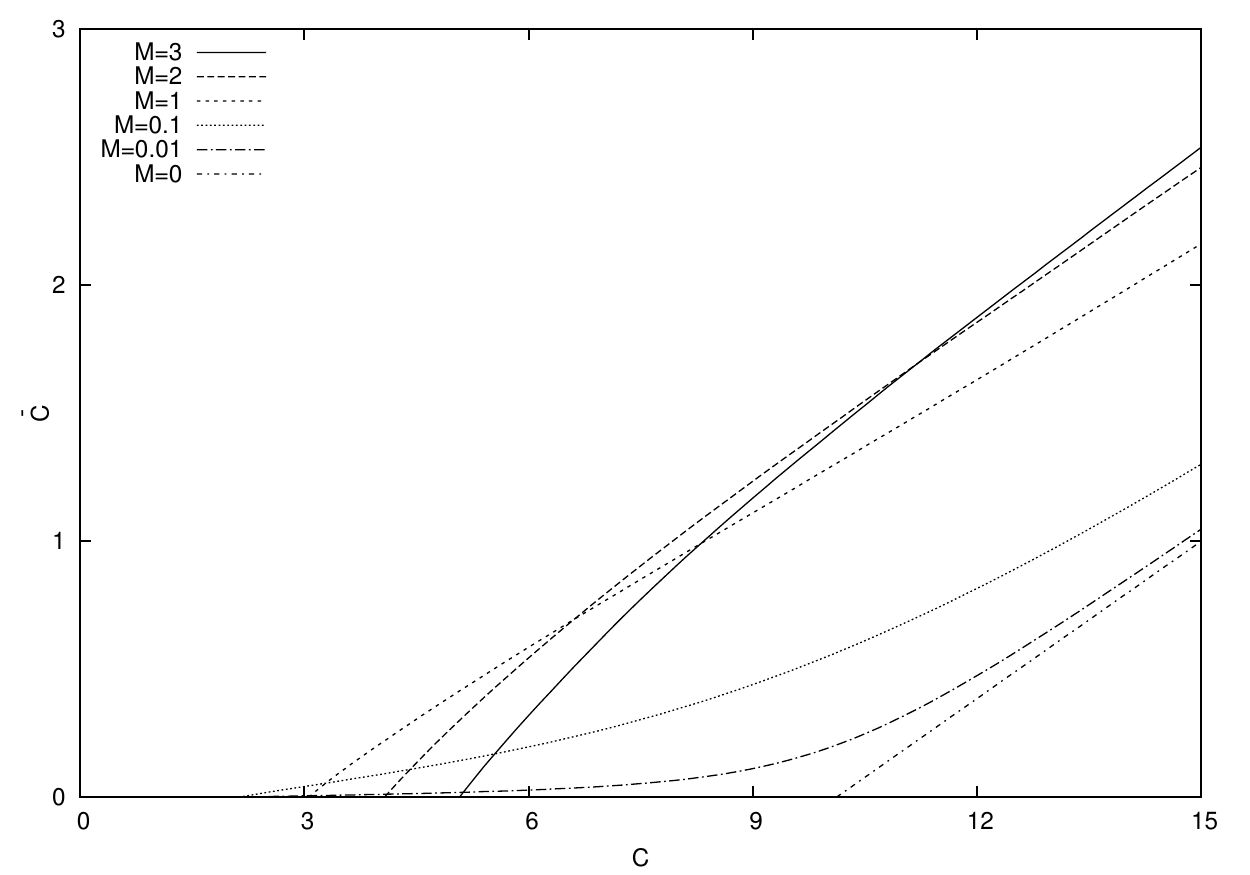}
\caption{Eigenvalue  $\bar{C}$  of the eigenvalue problem (\ref{psi}) as function of  the coexistence pressure $C$ for a non-degenerate
template competition scenario  with $S=1$ and  values
of the migration parameter $M$ as indicated in the figure. The transition point jumps from   $C_c = \pi^2 + 1/4$ for  $M=0$ to  
$C_c \approx 2$ for $M \to 0$.  }
\label{fig:3}
\end{figure}

\begin{figure}
\includegraphics{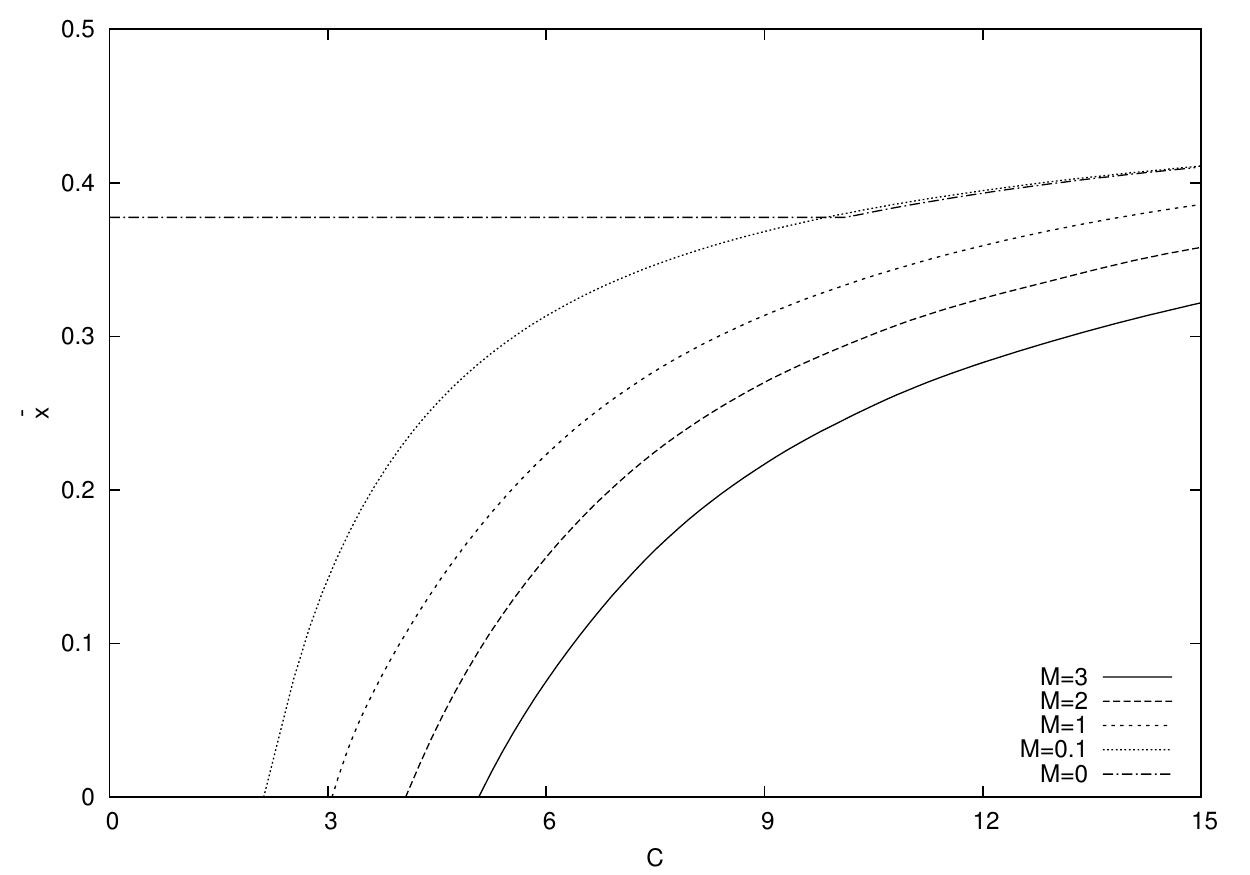}
\caption{Mean frequency of type 1 templates $\bar{x}$  as function of  the coexistence pressure  $C$ for $S=1$ and  
$M$ as indicated in the figure. For $M=0$, the segregating phase for $C < \pi^2 + S^2/4 \approx 10.12$ is non-ergodic
and the  result exhibited was obtained with the initial probability density $\phi \left ( x,0 \right ) = \delta \left ( x -1/2 \right )$ (see Appendix A). }
\label{fig:4}
\end{figure}

The scenario becomes more interesting when  the replication rates of the template types
 are allowed to differ, as illustrated in Figs.\ \ref{fig:3} and \ref{fig:4} for $S=1$. The first
noteworthy result exhibited in these figures is the appearance of a phase transition separating the homogeneous regime dominated by the
more efficient template type and  characterized by $\bar{x}=\bar{C} =0$, from the coexistence regime, $\bar{C} > 0$.
We note that  $\bar{x} > 0$ does not imply coexistence,
since this condition  holds true in the segregating phase that exists for $M=0$ and is characterized by an unbalanced mixture of
delta functions at the extremes $x=0$ and $x=1$.  Hence the eigenvalue $\bar{C}$ is the order parameter of our group selection 
diffusion model.
Fig.\ \ref{fig:4}   offers a better view of the  transition and highlights the
singular nature of the segregation phase for $M=0$.
Overall the effect of migration for $S>0$ is to hamper coexistence, as indicated by the need of a larger coexistence
pressure to establish the coexistence regime as $M$ increases. However, the transition from $M=0$  to an arbitrarily small
migration  value $M \to 0$ results  in a discontinuous jump  on the value of the minimal
coexistence
pressure needed to stabilize the coexistence phase (e.g., from $C \approx 10.12$ to $C \approx 2$ for $S=1$). 
As pointed out in Sec. \ref{sec:Steady}, this is so because the $M=0$ non-ergodic
segregating phase, characterized by the combination of delta functions $\phi = A _0 \delta \left ( x \right ) + A_1 \delta \left ( x - 1 \right )$,
with $A_0 + A_1 = 1$, is unstable to the effect of migration $M > 0$.   In this phase, $\bar{x} = A_1$ depends on the initial probability density 
(see Appendix A) and for $\phi \left ( x, 0 \right ) = \delta \left ( x - 1/2 \right )$ we find $A_1 = 1/\left [ 1 + \exp \left ( S/2 \right ) \right ]$ which is
depicted in Fig.\ \ref{fig:4}.
In the ergodic  phase (i.e., $C \geq \pi^2 + S^2/4$), however, the value  of $\bar{x}$ at $M=0$ is approached  smoothly in the limit 
$M \to 0$. The same is true for the order parameter $\bar{C}$  (see Fig.\ \ref{fig:3}), except that in this case the behavior is 
continuous for all values of $C$.

\subsection{The critical line}

The critical line separates the homogeneous from the coexistence regime. Since  at this line  $\bar{x}=\bar{C} = 0$, eq. (\ref{z})  
reduces to
\begin{equation}\label{zc}
\left  (1-x \right ) \left [ z_c' + z_c^2 \right ] -  \left  [ S \left (1-x \right ) + M \right ] z_c
+ C  \left  (1-x \right )  =  0  
\end{equation}
with
\begin{equation}\label{z_0}
z_c (0) \equiv z_0 = \lim_{\bar{C}, ~\bar{x} \to 0}  \frac{\bar{C}}{M \bar{x}}
\end{equation}
and
\begin{equation}\label{z_c1}
z_c (1) = 0 .
\end{equation} 
For fixed values of  the model parameters $S$, $C$ and $M$,  eq.\ (\ref{zc})  can be solved numerically 
by propagating the solution from $x=1$ to $x =0$ using  the Runge-Kutta algorithm. Thus,
given an arbitrary set of model parameters, eq.\ (\ref{zc})  has a unique solution under condition (\ref{z_c1}), which then
determines  $z_0$  univocally.  However,  since eq.\ (\ref{zc}) is valid  at the critical line  only we need another condition
to  constraint the values of the model parameters. Of course, this supplementary condition is  provided by  eq. (\ref{z_0}), which
reads
\begin{eqnarray}\label{critical}
z_0  & = & \frac{C}{M} \frac{\int_0^1 dx \exp \left ( - S x + y_c \right ) \left ( 1-x \right )^M}{\int_0^1 dx \exp \left ( - S x + y_c \right ) 
\left (1-x \right )^{M-1}} \nonumber \\
&  = & \frac{ C \int_0^1 dx \exp \left ( - S x + y_c \right ) \left ( 1-x \right )^M}{1 + \int_0^1 dx \exp \left ( - S x + y_c \right ) 
\left (1-x \right )^{M} \left ( -S + z_c  \right )},
\end{eqnarray}
where $y_c = \int_0^x z_c (\xi) d\xi$. The second line of this equation  is derived from the  first line by integration by parts and its
sole purpose is to emphasize the fact that $z_0$ is finite for $M \to 0$. The limits $\bar{x} \to 0$ and $\bar{C} \to 0$ were omitted
in eq. (\ref{critical}), so it is left implicit that this expression must be evaluated for values of  $S$, $C$ and $M$ at the
critical line. The critical line is then obtained by fixing $S$ and $M$ and adjusting  $C$ such that the value of  $z_c$ 
at the $x=0$ boundary of eq.\ (\ref{zc}) coincides with the value obtained using expression (\ref{critical}). 
This procedure is illustrated in  Appendix B for  the limit $M \to 0$,  where we can obtain  the analytical solution of 
eq.\ (\ref{zc}) as well as carry out explicitly the integrals in eq.\ (\ref{critical}).

\begin{figure}
\includegraphics{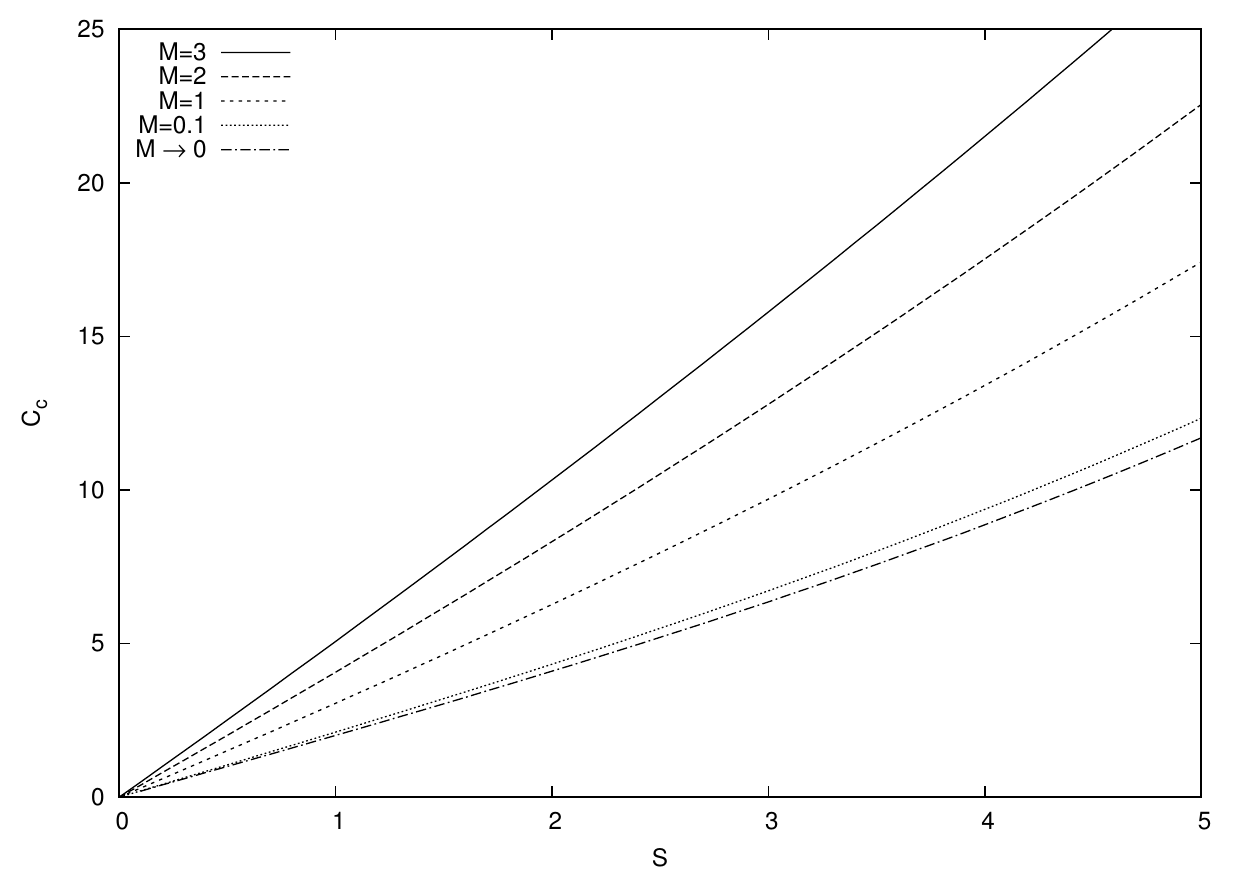}
\caption{Critical coexistence pressure $C_c$ as function of the selective advantage $S$ of type 2 templates.  For fixed $M$, the 
lines separate the homogeneous regime ($C \leq C_c$) where
the population is dominated by type 2 templates  from the coexistence regime
($C > C_c$) where both template types cohabit the same  protocell.
 }
\label{fig:5}
\end{figure}

The final outcome of the self-consistent iterative procedure described above is summarized in Fig.\ \ref{fig:5}. On the one hand, these
results support the conclusion that  for fixed $S>0$ increasing the migration rate $M$ hinders coexistence since
it is then  necessary to increase the coexistence pressure $C$  to guarantee the onset of the coexistence phase.
On the other hand,  a vanishingly small migration rate, represented by the curve $M \to 0$ in Fig.\ \ref{fig:5}, constitutes a
huge benefit to coexistence, as compared with the no-migration situation $M=0$ when the onset of the coexistence phase happens for 
$C> C_c = \pi^2 + S^2/4$ only \citep{Fontanari_13}.
The  reason is that for $M=0$ both template types are present in the  population but reside in distinct protocells, and so
a vanishingly small migration rate allows their meeting in a same protocell.

\subsection{Analytical approximation}

In the case $C$ and $S$ are small we can easily derive explicit expressions for the order parameter $\bar{C}$, as well as for
$\bar{x}$, and so obtain an analytical expression for the critical lines shown in Fig.\ \ref{fig:5}.  As pointed out, the assumption that
$C \ll 1 $ and $S \ll 1$ amounts to saying that these two selective pressures are small with respect to random drift and migration.
Considering the regular solution $\phi = \phi_r \left ( x \right )$ of eq.\ (\ref{stat1}), 
we begin by multiplying that equation by $x$ and then integrating over the interval $\left [ 0, 1 \right ]$,
yielding
\begin{equation}\label{ap1}
-S \int_0^1  \phi_r \left ( x \right )  x \left ( 1 - x \right )  dx + C \int_0^1  \phi_r \left ( x \right )  x^2 \left ( 1 - x \right )  dx -
\bar{x} ~C \int_0^1  \phi_r \left ( x \right )   x \left ( 1 - x \right )  dx =0.
\end{equation}
Next, to obtain results  that are correct to first order in  $S$ and $C$, we need only to replace $\phi_r$ by its
expression for $S=C=0$ [see eq.\ (\ref{phi0})], namely, the Beta distribution 
\begin{equation}\label{ap2}
\hat{\phi}_0 = \frac{x^{M\bar{x} -1} \left ( 1 - x \right )^{M\left ( 1 - \bar{x} \right ) -1}}{ B \left [ M \bar{x} , M \left ( 1 - \bar{x} \right )\right ]}
\end{equation}
where $B \left ( x,y \right )$ is the standard Beta function \citep{Abramowitz_72}. The final result is simply
\begin{equation}\label{ap3}
\bar{x} = \frac{1}{2} \left [ 1 -  \left ( M + 2 \right )  \frac{S}{C} \right ] ,
\end{equation}
from where we get $C_c = \left ( M + 2 \right )  S$ which agrees with the curves shown in Fig.\ \ref{fig:5} for small $S$.
For $S=0$ eq. (\ref{ap3}) yields $\bar{x} =1/2$ which is actually valid for all $C$ since 
$\phi_r \left ( x \right ) = \phi_r \left (1 - x \right )$ in this case.

Finally, to first order in $S$ and $C$ the order parameter $\bar{C}$ is given by
\begin{equation}\label{ap4}
\bar{C} = C \int_0^1 \hat{\phi}_0 \left ( x \right ) x \left ( 1 - x \right ) dx  = \frac{C M}{M+1} \bar{x} \left ( 1 - \bar{x} \right )
= \frac{CM}{4\left ( M+1 \right )} \left [ 1 - \left ( M+2 \right )^2 \frac{S^2}{C^2} \right ] ,
\end{equation}
which fits  very well  the curves  of Fig.\ \ref{fig:1} in the small $C$ regime, but fails to describe
the results of Fig.\ \ref{fig:3}  for $S=1$ since in that case the condition of small $S$  is not
satisfied.

\subsection{Discussion}\label{sec:Disc}

Here we address two issues that  were somewhat glossed over in the previous sections. The first issue
is the difference between the limit $M \to 0$ and the case $M=0$. From the physical perspective, that
difference is clear: in the absence of migration ($M=0$) there appears a segregation phase for $C < \pi^2 + S^2/4$ which is
unstable to the effect of a vanishingly small migration rate ($M \to 0$). However, from the mathematical 
perspective that difference is blurred by the fact that   the limit $M \to 0$ is obtained simply by setting $M=0$ in
our equations. The key point here is that  by writing the regular solution of  eq.\ (\ref{stat1}) in the form 
$ \phi_r \left ( x \right ) \propto \phi_0 \left ( x \right )  \psi \left ( x \right ) $ with  $\phi_0$  and $\psi$ 
given by eqs.\ (\ref{phi0}) and (\ref{psi}), respectively, we constrained the subsequent analysis to the region 
$M >0$ only, since  in that form $\phi_r$ is not normalizable for $M=0$. 
We note that for $M=0$ the regular solution of eq.\ (\ref{stat1}), which exists for  $C \geq \pi^2 + S^2/4$, 
is  finite at the extremes $x=0$ and $x=1$ and so it  is always  normalizable, as expected \citep{Fontanari_13}. Thus setting $M=0$  in 
eqs.\ (\ref{zc}) and (\ref{critical}) actually  means taking the limit $M \to 0$ of eq.\  (\ref{stat1}).

The second issue concerns the uniqueness of the eigenvalue $\bar{C}$ of  the second-order differential equation  for 
$\psi$, eq.\ (\ref{psi}). In fact,  if  there were no constraints on $\psi$ then there would be an infinity of admissible values 
for  the eigenvalue $\bar{C}$ as well as for the eigenfunctions $\psi$.  It is the condition that $\psi$ be positive and normalizable 
that reduces the acceptable solutions to a single one.  We note that by writing $\psi = e^y$ and solving numerically 
for  $y$ (see  eq.\ (\ref{eqy})) we have automatically restricted the numerical analysis to the valid regime $ \psi > 0$ only.

\section{Conclusion}\label{sec:Conc}
%

Contrary to the acrimony that has accompanied  the group selection accounts of  altruism
 and eusociality since  the 1960s \citep{Vero_62,Williams_66,Nowak_10,Rousset_11}, 
 group selection ideas have been mainstream in the prebiotic evolution  context \citep{Michod_83,Alves_01} since 
 there is  a consensus that the compartmentalization of  templates was an essential stage in the
process of molecular evolution \citep{Bresch_80,Eigen_80}. In addition,  compartmentalization offers a solution to
 the problem
of the coexistence between different templates \citep{Niesert_81,Silvestre_08}, which is the topic we address in this paper.
We should mention, however, that within the  context of the maintenance of cooperation the group selection or, more generally, 
the multilevel selection approach
has been applied to the study of the dynamics of cancer, which may be viewed as a result of the breakdown of cooperation 
between cells in the body \citep{Michor_04,Bellomo_08,Bellouquid_13}.

In this contribution we build on the  seminal paper by \citet{Kimura_83}, which  presented a diffusion model incorporating group
selection, and study a group selection pressure towards
the coexistence of two types of templates  that are differentiated by their replication rates. Our focus is on
the effect of template swapping (migration)  among protocells. This is
a key process within the modern prebiotic scenario, which is  based on the radical notion of an ancestral community
of cell lines lacking long-term genetic history and individuality, rather than of a single ancestral organism \citep{Woese_98}.

We find that the  progression of the template type that exhibits the selective advantage  at the individual level is greatly  promoted by  
migration, in the same manner that an antibiotic resistant gene spreads among a population comprising different bacterial species.
In that sense, migration hinders coexistence. Nevertheless, migration is very effective to counterweight the homogenizing effect 
of random drift (i.e., the fixation of a template type) so that in the degenerate case, where there is no selective advantage at the individual level, coexistence  is
the only possible outcome of the evolutionary  process. In addition,  even in the non-degenerate case, a small amount of template swapping
increases greatly the parameter range for which coexistence is stable in comparison with the case where there is no migration at all.
 
An interesting aspect of the diffusion model of  group selection  is the existence  of a continuous transition  between a homogeneous regime dominated by the
more efficient template type and a coexistence regime where the two template types cohabit a same protocell. The order
parameter that characterizes these regimes is the eigenvalue $\bar{C}$ of the eigenvalue problem (\ref{stat1}), whose eigenfunction is
the fraction of protocells with a given template composition at the the steady state. 
In particular, we find $\bar{C}>0$  in the coexistence regime,  and $\bar{C}=0$  in the homogeneous regime  
with $\bar{C}$ vanishing linearly with the distance to  the critical line that separates those regimes.

A simplifying feature of the model with migration is that the evolutionary dynamics is ergodic, i.e., the steady-state solution does not
depend on the details of the initial distribution of templates among the protocells, provided the two template types are present in
the population at the initial time. In fact, in the homogeneous phase  the more efficient template type fixates in all protocells with probability one,
whereas in the coexistence phase the distribution of template compositions inside the protocells are described univocally by the regular
solution  of eq.\ (\ref{stat1}).  The dynamics is non-ergodic  only in the  segregating phase  that appears for low  
coexistence pressure values  in the case migration is not allowed \citep{Fontanari_13}. For that case, we derive in  Appendix A
exact analytical expressions for the probability that one of the two template types fixates in a given protocell. Most interestingly,
this kind of local fixation  occurs both in the ergodic and in the non-ergodic phases of the model in the absence of migration
and so this model  offers a rare instance of subdivided population where the (local) fixation probabilities can  be calculated  exactly  
\citep{Slatkin_81,Blythe_07}.

To conclude, a word is in order about the stability of  the steady-state solutions of the non-linear (and non-local) partial differential equation
that determines the time evolution of the protocell population, eq.\ (\ref{eqphi}). On physical grounds one expects the existence of a coexistence regime for
large values of the coexistence group selection pressure  $C$ and so the stability of the steady-state regular solution 
$\phi = \phi_r \left ( x   \right)$, which
satisfies eq.\ (\ref{stat1}). In addition, in the absence of the coexistence pressure ($C=0$) the only steady-state solution is the homogeneous
one, i.e., $\phi = \delta \left ( x  - 1 \right) $. Whereas the regular solution exists  for $C > C_c \approx \left ( M + 2 \right ) S$ only, the homogeneous solution exists
for all $C \geq 0$ and so a possible instability of the regular solution at a finite value of  $C > C_c$ would shift the transition point as well as turn the
transition from continuous to discontinuous, in the sense that  the eigenvalue $\bar{C}$ would jump to zero at the new hypothetical transition point.
The analysis of the stability of the steady-state solutions by  techniques
such as the spectral theory in infinite dimensions \citep{Engel_00} is a most interesting and challenging enterprise that could reveal the influence of the
parameters $S$, $C$ and $M$  on the   relaxation time to  equilibrium as well as confirm the steady-state prediction of  the critical point  
$C_c$
separating the homogeneous and coexistence regimes. We hope our paper will motivate further studies on this research line.

 
\section*{Appendix A: Local fixation probability for the $M=0$  non-ergodic segregation regime}

As shown by \citet{Fontanari_13}, setting $M=0$ in eq.\ (\ref{stat1}) yields two possible steady-state solutions:
the solution corresponding to the ergodic coexistence phase, which is a combination of two Delta functions and a regular
function,
$\phi   \left ( x \right ) = A_0 \delta \left ( x \right ) + A_1 \delta \left ( x - 1\right ) + B \phi_r \left ( x \right ) $, with $A_0 + A_1 + B = 1$,
and the solution corresponding to the non-ergodic segregation phase, which is a combination of the two Delta functions,
$\phi   \left ( x \right ) = A_0 \delta \left ( x \right ) + A_1 \delta \left ( x - 1\right )$, with $A_0 + A_1 = 1$.
The non-ergodic regime, which is our focus here, occurs for $C < \pi^2 + S^2/4$.
Note that  in both regimes $A_1$ may be interpreted as the probability that 
the type 1 template fixates in a given protocell and  a similar interpretation holds for $A_0$ as well. However,  the result  $\bar{x} = A_1$,
which we used to draw the curve for $M=0$ in fig.\ \ref{fig:4}, holds in the
segregation regime only.
 In \citet{Fontanari_13} we have calculated the dependence of the weight $A_1$ on the
initial probability density $\phi \left ( x, 0 \right) $ for $S=0$ only, and  in this appendix we generalize that calculation for $S \geq 0$.

We begin by rewriting eq.\ (\ref{eqphi}) for $M=0$,
\begin{equation}\label{A1}
\frac{\partial}{\partial \tau} \phi \left (x,\tau \right ) = \frac{\partial^2}{\partial x^2} \left[ x \left (1-x \right )\phi \left (x, \tau \right ) \right]
+ S \frac{\partial}{\partial x} \left[ x \left (1 - x \right ) \, \phi \left (x,\tau \right ) \right]
+ \left  [ C x \left (1-x \right )-\bar{C} \left (\tau \right ) \right ] \phi  \left (x, \tau \right ) 
\end{equation}
and introducing the  abbreviation
 $\left \langle f  \left ( x \right ) \right \rangle_\tau = \int_0^1 f \left ( x \right ) \phi \left ( x , \tau \right ) dx$ for 
the expected value of  a regular function  $f \left ( x \right )$ at time $\tau$. Hence
\begin{eqnarray}\label{s_f1}
\frac{d}{d\tau} \left \langle f  \left ( x \right ) \right \rangle_\tau  & = & \left \langle x \left ( 1- x \right ) \frac{\partial^2 f (x)}{\partial x^2} \right \rangle_\tau
-S \left \langle x \left ( 1- x \right ) \frac{\partial f (x)}{\partial x} \right \rangle_\tau
  \nonumber \\
&  &  + C \left \langle x \left ( 1- x \right ) f  \left ( x \right ) \right \rangle_\tau 
- \bar{C} \left ( \tau \right ) \left \langle f  \left ( x \right ) \right \rangle_\tau
\end{eqnarray}
with  $\bar{C} \left ( \tau \right )  = C \left \langle x \left ( 1- x \right ) \right \rangle_\tau$.   The idea is to choose
a function $f(x)$ such that the first three  terms of the right hand side of eq.\ (\ref{s_f1})  cancel  out.  This choice
depends on the value of the parameter  $\Gamma  \equiv C -  S^2/4$ as discussed next. We note that $\Gamma < \pi^2$
in the non-ergodic regime.

\paragraph{Region $0 < \Gamma < \pi^2$.}

In this region we choose $f \left ( x \right ) = e^{S x/2} \sin \left ( \sqrt{\Gamma} x + \theta \right) $ where
$\theta$ is an arbitrary constant. Then eq.\ (\ref{s_f1}) rewrites
\begin{equation}
\frac{d}{d\tau} \left \langle  e^{S x/2} \sin \left ( \sqrt{\Gamma} x + \theta \right) \right \rangle_\tau
= - \bar{C} \left ( \tau \right ) \left \langle  e^{S x/2} \sin \left ( \sqrt{\Gamma} x + \theta \right) \right \rangle_\tau
\end{equation}
which has the formal solution 
\begin{equation}
\frac{\left \langle  e^{S x/2} \sin \left ( \sqrt{\Gamma} x + \theta \right) \right \rangle_\tau}{\left \langle  e^{S x/2} \sin \left ( \sqrt{\Gamma} x + \theta \right) \right \rangle_0} = \exp \left [ - \int_0^\tau \bar{C} \left ( \eta \right ) d \eta \right ] .
\end{equation}
As the right hand side of this equation does not depend on $\theta$, neither does the ratio in its left hand side. Hence, equating
the ratios evaluated at $\theta = 0$ and $\theta = \pi/2 - \sqrt{\Gamma}/2$ yields
\begin{equation}
\frac{\left \langle  e^{S x/2} \sin \left ( \sqrt{\Gamma} x \right) \right \rangle_\tau} 
{\left \langle  e^{S x/2} \cos \left [ \sqrt{\Gamma} \left ( x - 1/2 \right ) \right] \right \rangle_\tau}
= \frac{\left \langle  e^{S x/2} \sin \left ( \sqrt{\Gamma} x  \right) \right \rangle_0} 
{\left \langle  e^{S x/2} \cos \left [ \sqrt{\Gamma} \left ( x - 1/2 \right ) \right] \right \rangle_0} .
\end{equation}
In the limit $\tau \to \infty$ we have 
\begin{equation}
 \left \langle  e^{S x/2} \sin \left ( \sqrt{\Gamma} x \right) \right \rangle_\infty =
A_1 e^{S/2} \sin \left ( \sqrt{\Gamma}  \right) 
\end{equation}
 and
\begin{eqnarray}
\left \langle  e^{S x/2} \cos \left [ \sqrt{\Gamma} \left ( x - 1/2 \right ) \right] \right \rangle_\infty & = &
\left ( A_0 + A_1 e^{S/2} \right ) \cos \left (\sqrt{\Gamma}/2 \right ) \nonumber \\
& = & 
\left [ 1 + A_1\left ( e^{S/2} - 1 \right )  \right ] \cos \left (\sqrt{\Gamma}/2 \right ) 
\end{eqnarray}
which leads to
\begin{equation}
A_1 = \frac{1}{1 +  e^{S/2} \left ( \Xi_0 -1 \right)}
\end{equation}
where
\begin{equation}
\Xi_0 = 2 \sin \left (\sqrt{\Gamma}/2 \right ) \frac {\left \langle  e^{S x/2} \cos \left [ \sqrt{\Gamma} \left ( x - 1/2 \right ) \right] \right \rangle_0}
{\left \langle  e^{S x/2} \sin \left ( \sqrt{\Gamma} x  \right) \right \rangle_0} .
\end{equation}
In the limit  $\Gamma \to  \pi^2$, we have $\Xi_0 \to 2$ regardless of  the initial probability density $\phi \left ( x, 0 \right )$ and so
$A_1 \to A_1^c =  1/\left (1 +  e^{S/2}  \right )$. In addition, for the initial probability density $\phi \left ( x, 0 \right ) = \delta \left ( x - 1/2 \right )$ used to
calculate $\bar{x}$ at $M=0$ in Fig.\ \ref{fig:4}, the dependence on $\Gamma$ (and hence on $C$) disappears and so
$A_1 = A_1^c$.

\paragraph{Region $-S^2/4 < \Gamma < 0$.}

In this region the choice $f \left ( x \right ) = e^{S x/2} \left ( e^{ux}  + \theta e^{-ux} \right ) $ with 
$u = \sqrt{-\Gamma}$ and $\theta$ arbitrary leads to the canceling
of the  first three  terms of the right hand side of eq.\ (\ref{s_f1}) yielding
\begin{equation}
\frac{\left \langle  e^{S x/2} \left ( e^{ux}  + \theta e^{-ux} \right )  \right \rangle_\tau}{\left \langle  e^{S x/2} \left ( e^{ux}  + \theta e^{-ux} \right ) \right \rangle_0} = \exp \left [ - \int_0^\tau \bar{C} \left ( \eta \right ) d \eta \right ] .
\end{equation}
The same argument used in the analysis of the  $\Gamma > 0$ region allows us to equate the ratio that appear in the left hand side of this
equation for $\theta = -1$ and $\theta =0$,
\begin{equation}
\frac{\left \langle  e^{S x/2} \sinh \left ( u x \right) \right \rangle_\tau} 
{\left \langle  e^{S x/2 +ux}  \right \rangle_\tau}
= \frac{\left \langle  e^{S x/2} \sinh \left ( u x \right) \right \rangle_0} 
{\left \langle  e^{S x/2 + ux} \right \rangle_0} .
\end{equation}
Finally, taking the limit $\tau \to \infty$ yields
\begin{equation}
A_1 = \frac{1}{1 +  e^{S/2} \left [ \Omega_0\sinh \left ( u \right ) -e^u  \right]}
\end{equation}
where
\begin{equation}
\Omega_0 = \frac {\left \langle  e^{S x/2 + ux } \right \rangle_0}
{\left \langle  e^{S x/2} \sinh \left ( u x  \right) \right \rangle_0} .
\end{equation}
By taking the limit $u \to 0$ we can easily verify that  $A_1$ is continuous at the boundary of
the two regions.  In addition, in the limit $C \to 0$, i.e., $ u \to S/2$ we recover the classical  formula
for the fixation of an allele with selective disadvantage $S$ \citep{Crow_70},
\begin{equation}
A_1 = \frac{\left \langle e^{Sx} \right \rangle_0 - 1}{e^S - 1 }.
\end{equation}
Similarly to our finding in the analysis of the previous region,  the initial probability density $\phi \left ( x, 0 \right ) = \delta \left ( x - 1/2 \right )$
yields $A_1 = 1/\left (1 +  e^{S/2}  \right )$ regardless of the value of $C$, as shown in Fig.\ \ref{fig:4}.

\section*{Appendix B: Critical line for the limit $M \to 0$}

Setting  $M = 0$ in eq.\ (\ref{zc})  yields
\begin{equation}\label{zcq}
z'_c + z^2_c - S z_c +C=0
\end{equation}
with the condition 
\begin{equation}\label{zc1}
z_c \left ( 1 \right )=0 .
\end{equation}
In the region  $C- S^2 /4 > 0$  its solution is
\begin{equation}\label{zcx}
z_c \left (x \right )= \frac{S}{2} -\gamma \tan \left ( \gamma x + \theta \right )
\end{equation}
where $\gamma=  \sqrt{C- S^2 /4}$ and $\theta = \theta \left (\gamma, S \right )$ 
is fixed by condition (\ref{zc1}) as 
\begin{equation}\label{zcf}
\frac{S}{2} -\gamma \tan \left ( \gamma + \theta \right )=0 .
\end{equation}
We note that the critical value   $C_c \left  (S \right )$  is in the region $C- S^2 /4 > 0$ (see Fig.\ \ref{fig:5}). 
To evaluate eq.\ (\ref{critical}) we use the equality
\begin{equation}\label{eyc}
\exp \left ( -Sx + y_c \right ) = 
\frac{\cos \left( \gamma x + \theta \right )}{\cos \left (\theta \right )}
 \; \exp \left ( -S x/2 \right ).
\end{equation}
which follows directly from the definition $y_c \left ( x \right ) = \int_0^x z_c \left ( \xi \right ) d \xi$ with 
$z_c$ given by (\ref{zcx}). Now the integrals in eq.\ (\ref{critical}) can be readily evaluated yielding
\begin{equation}\label{tli}
\frac{S}{2} -\gamma \tan \left ( \theta \right )=
\frac{
\gamma \left [ e^{-\frac{S}{2}}\sin \left( \gamma+ \theta \right ) -\sin \left ( \theta \right ) \right ]
- \frac{S}{2} \left [ e^{-\frac{S}{2}}\cos \left (\gamma+ \theta \right )-\cos \left ( \theta \right) \right ]
}
{e^{-\frac{S}{2}}\cos \left (\gamma+ \theta \right )} .
\end{equation}
This equation can be further simplified using the equalities 
 $\sin \left ( \gamma + \theta \right )= S/ \left ( 2\sqrt{C} \right )$
and $\cos \left (\gamma + \theta \right )=\gamma/\sqrt{C}$ that follow from eq.\ (\ref{zcf}).
The final result is simply
\begin{equation}\label{tl2}
\gamma = \sqrt{C} e^{\frac{S}{2}}\cos \left ( \theta \right ) .
\end{equation}
Finally, we rewrite eq.\  (\ref{zcf}) as
\begin{equation}\label{tl3}
\theta = \arctan\left( \frac{S}{2\gamma}\right) - \gamma
\end{equation}
in order to make clear that eq.\ (\ref{tl2}) yields a relation
$ C = C_c  \left ( S \right )$,
which is  the critical line $M \to 0$ depicted in Fig.\ \ref{fig:5}.

\begin{acknowledgements}
The research of J.F.F. was supported in part by Conselho Nacional de Desenvolvimento
Cient\'{\i}fico e Tecnol\'ogico (CNPq) and the research of M.S. was 
partially supported by PRIN 2009 protocollo n.2009TA2595.02.
\end{acknowledgements}



\end{document}